# Exciton-polariton topological insulator

S. Klembt[1,+,*], T.H. Harder[1,+], O.A. Egorov[1], K. Winkler[1], R. Ge[2], M.A. Bandres[3], M. Emmerling[1], L. Worschech[1], T.C.H. Liew[2], M. Segev[3], C. Schneider[1] and S. Höfling[1,4]

[1]Technische Physik and Wilhelm-Conrad-Röntgen-Research Center for Complex Material Systems, Universität Würzburg, Am Hubland, 97074 Würzburg, Germany
[2]Division of Physics and Applied Physics, School of Physical and Mathematical Sciences, Nanyang Technological University, 21 Nanyang Link, 637371, Singapore
[3]Physics Department and Solid State Institute, Technion, Haifa 32000, Israel
[4]SUPA, School of Physics and Astronomy, University of St. Andrews, KY 16 9SS, United Kingdom
[+]These authors contributed equally to this work
*sebastian.klembt@physik.uni-wuerzburg.de

**Topological insulators are a striking example of materials in which topological invariants are manifested in robustness against perturbations [1,2]. Their most prominent feature is the emergence of topological edge states with reduced dimension at the boundary between areas with distinct topological invariants. The observable physical effect is unidirectional robust transport, unaffected by defects or disorder. Topological insulators were originally observed in the integer quantum Hall effect [3], and subsequently suggested [4-6] and observed [7] even in the absence of magnetic field. These were fermionic systems of correlated electrons. However, during the past decade the concepts of topological physics have been introduced into numerous fields beyond condensed matter, ranging from microwaves [8,9] and photonic systems [10-12] to cold atoms [13,14], acoustics [15,16] and even mechanics [17,18]. Recently, topological insulators were proposed [19-21] in exciton-polariton systems organized as honeycomb (graphene-like) lattices, under the influence of a magnetic field. Topological phenomena in polaritons are fundamentally different from all topological effects demonstrated experimentally thus far: exciton-polaritons are part-light part-matter**

**quasiparticles emerging from the strong coupling of quantum well excitons and cavity photons [22]. Here, we demonstrate experimentally the first exciton-polariton topological insulator. This constitutes the first symbiotic light-matter topological insulators. Our polariton lattice is excited non-resonantly, and the chiral topological polariton edge mode is populated by a polariton condensation mechanism. We use scanning imaging techniques in real-space and in Fourier-space to measure photoluminescence, and demonstrate that the topological edge mode avoids defects, and that the propagation direction of the mode can be reversed by inverting the applied magnetic field. Our exciton-polariton topological insulator paves the way for a variety of new topological phenomena, as they involve light-matter interaction, gain, and perhaps most importantly – exciton-polaritons interact with one another as a nonlinear many-body system.**

Microcavity exciton-polaritons (polaritons) are composite bosons originating from the strong coupling of quantum well excitons to microcavity photons. While the excitonic fraction provides a strong non-linearity, the photonic part results in a low effective mass, allowing the formation of a driven-dissipative Bose-Einstein condensate [23,24] and a superfluid phase [25], making polaritons being referred to as „quantum fluids of light" [26]. For the epitaxially well-controlled III-V semiconductor material system, a variety of techniques are available to micropattern such cavities in order to precisely engineer the potential landscapes of polaritons [27]. With the recent advances of bringing topological effects to the realms of photonics [8-12,28], several avenues to realize topological edge propagation with polaritons have been suggested [19-21], with honeycomb geometries ("artificial graphene" [29,30]) being of particular interest to realize a C=2 Chern band insulator [21]. Indeed, polariton honeycomb lattices have been found to support Dirac-cone dispersions [31] as well as edge modes [32,33] inherited from their graphene origin [34,35].

Here, we make the next step and realize a topological Chern insulator in the exciton-polariton system, based on the system proposed in [19-21]: a honeycomb potential landscape with its time-reversal symmetry broken by an applied magnetic field.

Let us first introduce the underlying physics and show its features in numerical simulations with a realistic set of sample parameters. The experiments display reliable condensation of polaritons at the vicinity of the Dirac cones and the creation of topological edge modes enabled by applying a magnetic field. Hence, we consider the injection of polaritons into the topological gap and focus on the inherent properties of the chiral edge modes.

A general schematic of the experiment is presented in Fig. 1a, while Figs. 1b and c depict the calculated dispersion relation of the honeycomb structure in the direction $\Gamma \to K$, connecting two Dirac cones ($K$ and $K'$), without and with applied magnetic field, respectively. The effective spin-orbit coupling of polaritons, induced by TE-TM mode splitting, breaks the polarization-related symmetry and thus each Dirac point transforms into four inverted parabolas [21]. While the spin-orbit interaction is extremely small in real graphene [36], "polariton graphene" offers the possibility to make the effective spit-orbit coupling sufficiently large to open a sizeable gap in a magnetic field. Without a magnetic field, the two central parabolas touch each other at the Dirac cones of the underlying honeycomb lattice (Fig. 1b). The degeneracy between the states in the crossing points can be lifted in the presence of a magnetic field and a finite Zeeman splitting. As a consequence, an energy gap forms in the vicinity of the Dirac cones (Fig. 1c). It is worth mentioning that the Dirac cones $K$ and $K'$ are not equivalent. At the $K$ ($K'$) point, the "valence" band is formed from the $B$ ($A$) pillars and the "conduction" band is formed from the $A$ ($B$) pillars [21]. The reversed order of the bands in the basis of the sublattices signifies that the gap is topologically non-trivial [1].

The interplay of an external magnetic field and the effective spin-orbit coupling (TE-TM mode splitting) [37] results in non-zero Berry connections around the $K$ and $K'$ points, contributing to the total band Chern number $C = \pm 2$ [20,21]. As a consequence, the honeycomb structure supports one-way propagating edge states for the energies (eigenfrequencies) within the topological gap. Fig. 1c demonstrates the results of a band structure calculation combined with the dispersion of the edge states localized at the zigzag edge of the honeycomb structure. The propagation direction of these edge states is related to the direction of the external magnetic field: the polariton edge current is either left moving (L, yellow) or right moving (R, red), depicted in Fig. 1a, depending on the sign of the magnetic field. Fig. 1e depicts a calculation of the corresponding edge mode.

To illustrate the existence and robustness of the topologically-protected one-way edge states, we simulate the evolution of a wavepacket excited locally (Figs. 1f, g (red circle)) at the zigzag edge of the honeycomb structure. Figs. 1f and g show that the launched wave packet starts to propagate left and right along the edge, respectively, depending on the polarity of the magnetic field, whereas in the absence of magnetic field, the launched wave packet remains at the excitation point. Moreover, the wave packet flips its propagation direction when the sign of the external magnetic field and thus the sign of the Zeeman splitting is reversed (see Figs. 1f, g). Note that the overall intensity decreases with propagation, since the model takes into account a realistic polariton lifetime of ~35 ps (for more details see Supplementary Information S1). Furthermore, the Supplementary Information S2 shows that the chiral edge mode propagates along a 90° corner and is able to bypass a point-like defect.

Having established the features we expect to observe in an exciton-polariton topological insulator, we describe the experimental platform. To realize a polariton honeycomb potential lattice, we fabricate a planar microcavity containing three In$_{0.04}$Ga$_{0.96}$As quantum wells (QWs) in a λ-cavity, sandwiched between two GaAs/AlGaAs distributed Bragg reflectors (DBRs) with 30 (35.5) top (bottom) mirror pairs (Fig. 2a). Subsequently, we use electron beam lithography to define the honeycomb lattice, employing micropillars with a diameter of 2.0 µm and a pillar-to-pillar overlap of $v = a/d = 0.85$, where $a$ denotes the center-to-center distance between neighboring pillars. Finally, the upper DBR is etched in such a way that only two DBR pairs of the top DBR remain, so as not to damage the active QW region (Figs. 2b, c) (see Methods Section for more details). When these sites are arranged in a two-dimensional lattice, the discrete pillar modes hybridize due to their proximity to one another and form a polaritonic band structure, analogous to the electronic band structure of graphene. The honeycomb lattice is characterized by a two-element base in real-space (Fig. 2d) and six degenerate K- and K'-points supporting Dirac cones in the first Brillouin zone (Fig. 2e), as well known from graphene. Fig. 2f depicts characterization of the polariton honeycomb lattice in the linear regime using non-resonant laser excitation. The Fourier-space energy-resolved photoluminescence of the investigated lattice is imaged in $k_y$-direction, and scanned in $k_x$-direction. The blue data points are fitted to the measured dispersions, accurately revealing the six Dirac cones at the K- and K'-points. The results of the corresponding tight-binding model (see Supplementary Information S3) are plotted in red and yellow, agreeing very well with the experimental data.

Next, we describe the experiments, conducted on the honeycomb exciton-polariton lattice under an external magnetic field, aiming to find topologically non-trivial edge states. The lattice used for these experiments is chosen based on TE-TM splitting, Zeeman splitting and linewidth properties, summarized in the Methods Section. Supplementary Information S4 presents a top-view microscopy image of the zigzag-edge of this lattice including an artificial defect, i.e. an unoccupied lattice site, which is used later on to study robustness. To be able to observe a band gap that opens at the Dirac points, the Zeeman-splitting as well as the TE-TM-splitting at the wavevector and energy of the Dirac points, need to be sufficiently large compared to the photoluminescence (PL) linewidth. This implies that the polaritons need to have a sufficient excitonic part. Therefore, we select a lattice at a moderate negative detuning of $\delta = -11.5$ meV. Furthermore, to increase the chances of being able to measure a band gap, a large slope of the dispersion at the Dirac points is advantageous.

In order to assess the size of the bandgap, we apply polarization resolved spectroscopy, making use of a λ/4 polarization series at an external magnetic field of $B = 0$T and 5T. A band gap of $E_g = (108 \pm 32)$ µeV at 5 T is evaluated (see Supplementary Information S5). At an external magnetic field of $B = +5$ T, the Hopfield coefficients at the Dirac points return a photonic fraction of $|C|^2 = 0.96$ and an excitonic fraction of $|X|^2 = 0.04$. Furthermore, the TE-TM-splitting at the wavevector of the Dirac point (i.e. $|k_D| \approx 0.77\frac{\pi}{a}$) can be estimated to be 400 µeV for the photons, resulting in an effective TE-TM splitting for the exciton-polarions $\simeq 384$ µeV, ($\beta_{eff} \simeq 263$ µeVµm$^2$). The Zeeman-splitting of the excitonic mode is determined to be ~540 µeV, leading to $\Delta_{eff} \simeq 22$ µeV at the Dirac point (see Supplementary Information S1 and S6). As the TE-TM-splitting is considerably larger than the Zeeman-splitting in the lattice studied here, the experimentally determined band gap that opens due to the magnetic field is of $\Delta_g = (108 \pm 32)$ µeV with $\Delta_{eff} < \Delta_g < \beta_{eff}$, which, as we show later on, is reasonable for observing the topological features of our lattice and compares well with a gap size of about 0.1 meV found for realistic system parameters in [20]. After having established that a band gap opens at the Dirac points under the influence of an external magnetic field, we perform non-resonant excitation, under conditions allowing for polariton condensation into a chiral edge state within this band gap. Scans of the PL signal in real space as well as Fourier space are performed on the polariton lattice at external magnetic fields of $B = -5T, 0T$ and $+5T$. The sample is excited on the zigzag-edge with a pulsed and chopped large laser spot with a diameter of 40 µm at a wavelength of 792 nm, tuned to the first stop band minimum (see Fig. 1a). In Fig. 3a the linear polariton dispersion in $K' \rightarrow \Gamma \rightarrow K$ direction at $B = 0T$ and (b) at $B = +5T$ is displayed. The Dirac-cone dispersion is clearly visible ($P \sim 0.1$ mW). At a threshold power of $P_{thr} = 1.8$ mW, we observe a nonlinear increase of the output power as well as a sudden decrease in linewidth (see Supplementary Information S7), establishing that a polariton condensate has formed at the K,K'-points at around $k_y \approx \pm 0.77 \frac{\pi}{a}$, as displayed in Fig. 3c. We now perform mode tomography, scanning the real space (x,y) landscape and measuring the energy E_PL. Fig. 3d shows the spectrum perpendicular to the zigzag edge, taken at the position indicated by the solid white line in Fig. 3g. The dashed white line depicts the edge of the sample. Remarkably, besides a S-band condensate throughout the excited structure at $E_S = 1.4674 \ eV$, we observe an edge mode: a region of high intensity residing only at the outermost row of lattice sites, at an energy of $1.4678 \ eV$ (white ellipse).

Fig. 3e shows the image of the intensity pattern integrated over the energetic range of the trivial S-band mode centered at 1.4674 eV. Clearly, the condensate is relatively homogeneous over a large fraction of the lattice. The white overlaid lattice geometry and the microscopy image insets illustrate the position of the edge of the sample. Now, by changing the energy of the mode tomography to an energy $E_{edge} = 1.4678 \ eV$, within the topological gap (under a magnetic field of +5T), the existence of edge states becomes unequivocal (at x≈8µm in Fig. 3g). The PL at this energy originates predominantly from the outermost row of lattice sites with almost no emission detected from the bulk of the lattice. The mode is in excellent agreement with the Bloch mode calculations in Figs. 1e, f, g. In addition, the theoretical description within a Ginzburg-Landau-based model confirms that indeed polariton condensation into the edge mode occurs (see

Supplementary Information S8). We now move on to study the robustness of the topological edge state. We do that by installing an artificial defect into the lattice at $y = 15$ µm (see white circle in Figs. 3e-g). The defect is formed by leaving one of the sites on the zigzag edge of the honeycomb lattice unoccupied (a missing pillar; see Supplementary Information S2 and S4). Normally, such a strong defect would cause scattering into the bulk, but here (see Figs. 3f, g) such scattering is suppressed indicating that the transport of the topological edge state is immune to such defects. In addition, we perform the same measurement at the corner position of the sample, which again highlights the topological protection of the edge state. When plotting the energy of the topological edge mode $E_{edge} = 1.4678\ eV$ in Fig. 3f, one clearly observes that the mode extends around the corner from the zigzag into the armchair configuration, without any sign of backscattering or bulk scattering. The measurements at $B = -5$T show very similar behavior, but the transport is in the opposite direction. Remarkably, when the magnetic field is absent ($B = 0$T), the edge mode vanishes completely (see Supplementary Information S10). The observation of the edge mode around the corner and its existence at the armchair edge (where edge states cannot exist in the absence of magnetic field) is not expected for trivial edge modes [34]. These observations therefore prove unequivocally that the edge mode we observe is indeed topological and is endowed with topological protection.

To get further insight into the nature of these edge states, we analyze hyperspectral images $\left((k_x, k_y)\ vs.\ E_{PL}\right)$ to identify the dominant propagation direction, with and without magnetic field. The results are displayed in Fig. 4. While in real-space the modes can be clearly separated in energy, the integration over Fourier-space results in the topological edge mode and the trivial bulk modes to overlay on top of one another. Figs. 4a, b depict the integrated intensities of the full S-band condensate (1.467 eV − 1.468 eV) including the energies associated with the edge state, for experiments at $B = +5\text{T}$ and $-5\text{T}$, respectively. To analyze the directionality of polariton transport, the maximum peak intensities at the two maxima at $k_x \approx 0$ and $k_y \approx \pm 0.77$ (π/a) are extracted, by identifying the central pixel of the peak and averaging the intensity of a region of 3 × 3 pixels centered around this position. The vertical axis of Fig. 4c shows the ratio of the luminescence travelling in one direction ($+k_y$) to the opposite direction ($-k_y$). Deviation of this quantity from unity is an essential characteristic of a chiral state and an opposite deviation should appear for opposite applied magnetic field. The corresponding intensity ratios are plotted in blue and show a clear directionality when an external magnetic field is applied. The transport changes its predominant direction along the edge when the direction of the magnetic field is inverted. This observation supports the interpretation of the edge-states being a result of a topologically non-trivial band gap, with the edge mode contributing to the chirality along the edge. Reversing the magnetic field reverses the slope of the dispersion curve of the topological edge mode, which is physically manifested in reversing the group velocity. On the other hand, we find no systematic directionality for the peaks at $k_x \approx -0.7 \left(\frac{\pi}{a}\right)$ in Figs. 4a, b, which implies that these arise solely from the bulk condensate.

The experimental results depicted in Figs. 3 and 4 prove, unequivocally, the observation of an exciton-polariton topological Chern insulator. The application of a magnetic field on the honeycomb lattice opens up a topological bandgap with topological edge states supporting unidirectional transport whose propagation direction is determined by the field polarity. The lack of scattering from an artificial defect manifests the robustness of the topological edge mode. Furthermore, the observation of the edge mode extending around the corner and at the armchair termination without bulk scattering is a distinct feature of the topological edge mode. Our results lead the way to efficient light trapping and topologically protected propagation of exciton-polariton condensates in a well-developed semiconductor platform, where also electrical driving can be envisaged [38,39]. We now aim at further exploring the topological lasing aspect of these experiments, by comparing topological edge mode lasing to lasing from a trivial edge mode in e.g. a Semenoff insulator [40] that could be realized employing this platform. Such experiments would also link our exciton-polariton platform to the recently observed topological insulator laser [41-44]. Altogether, this work paves the way towards new topological polaritonic devices with unique properties and functionalities, based on topological polariton edge transport [45].

**Acknowledgements**

The authors thank R. Thomale for fruitful discussions. S.K. acknowledges the European Commission for the H2020 Marie Skłodowska-Curie Actions (MSCA) fellowship (Topopolis). S.K., S.H. and M.S. are grateful for financial support by the JMU-Technion seed money program. S.H. also acknowledges support by the EPSRC "Hybrid Polaritonics" Grant (EP/M025330/1). The Würzburg group acknowledges support by the ImPACT Program, Japan Science and Technology Agency and the State of Bavaria. T.C.H.L. and R. G. were supported by the Ministry of Education (Singapore) Grant No. 2017-T2-1-001.

**Authors contributions**

S.K., M.S., C.S. and S.H. initiated the study and guided the work. S.K., T.H., K.W., M.E. and S.H. designed and fabricated the device. S.K. and T.H. performed optical measurements. S.K., T.H., O.A.E. and C.S. analyzed and interpreted the experimental data. O.A.E., R.G., T.C.H. L., M.B. and M.S. developed the theory. S.K., T.H., O.A.E., T.C.H. L., C. S., M.S. and S. H. wrote the manuscript, with input from all coauthors.

**Additional information**

**Competing financial interests:** The authors declare no competing financial interests.

**Correspondence and requests** for materials should be addressed to Sebastian Klembt (sebastian.klembt@physik.uni-wuerzburg.de) and Sven Höfling (sven.hoefling@physik.uni-wuerzburg.de)

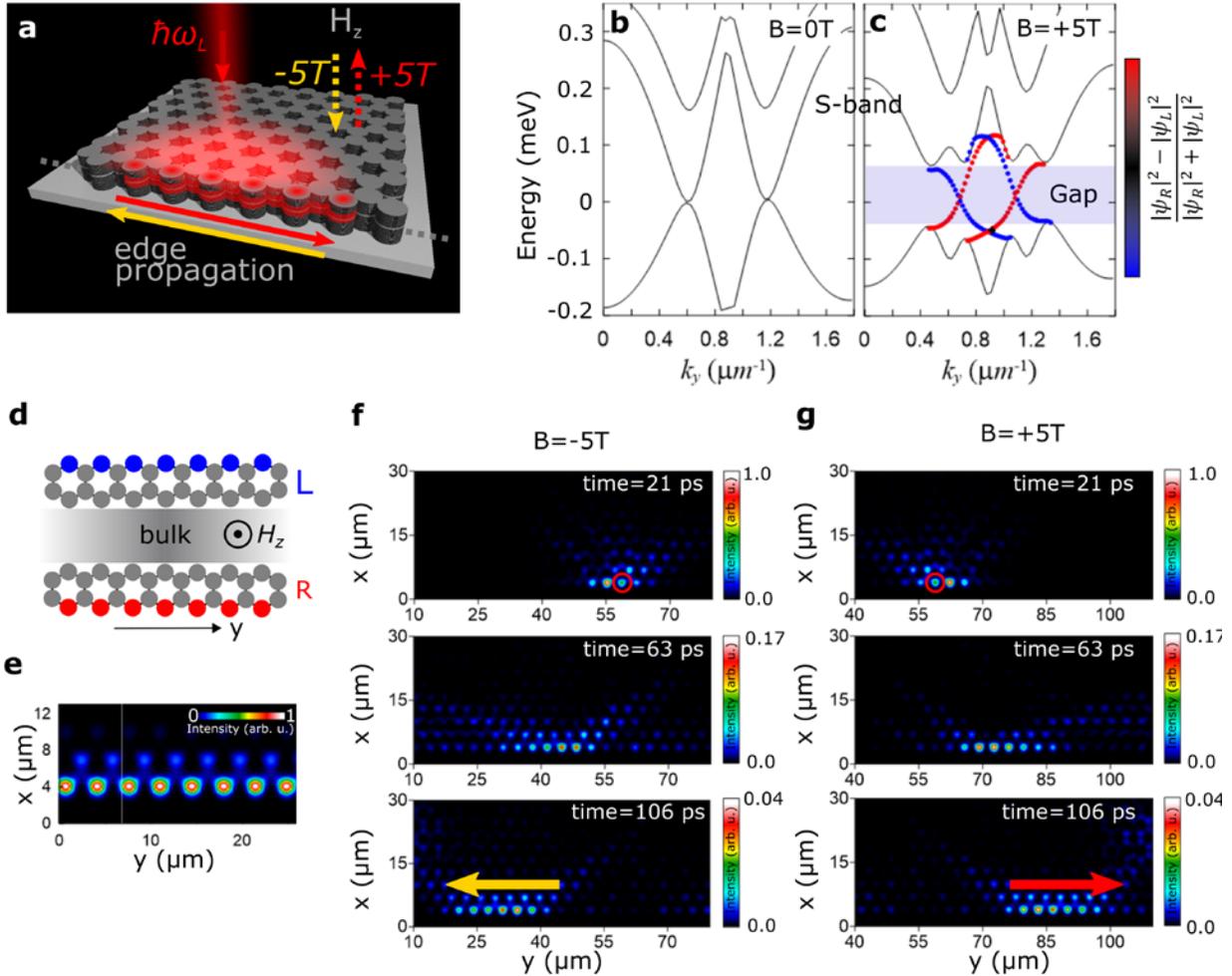

**Figure 1 | Experimental scheme and theoretically calculated Bloch mode and dynamics calculations of topological edge mode.** (**a**) Schematic of the experiment, exciting non-resonantly with a large diameter laser spot chiral topological edge modes (left moving, yellow and right moving, red) appearing in a gap opened by a combination of photonic spin-orbit coupling (TE-TM splitting) and applied magnetic field. (**b, c**) Trivial and topological band structures of the polariton honeycomb lattice for zero Zeeman splitting (b) and with Zeeman splitting $\Delta_{eff}=$ 0.2 meV, induced by the external magnetic field (**c**). Effective spin orbit coupling is induced by the TE-TM splitting of the photonic component $\beta_{eff} =$ 0.1 meVμm$^2$ . The one-way topological edge modes are presented by the red and blue lines within the topological gap. (**d**) Schematic and (**e**) calculated intensity profiles of the edge modes. (**f, g**) Calculated propagation dynamics of edge modes injected coherently into the topological. Here, (**f**) shows the left moving time-sequence (yellow) for the negative value of the Zeeman splitting $\Delta_B= -0.8$ meV and (**g**) shows the right moving propagation (red) for the positive value splitting $\Delta_B= +0.8$ meV. A linearly-polarized narrow seeding coherent beam injects both polarization components into the pillar marked by the red circle.

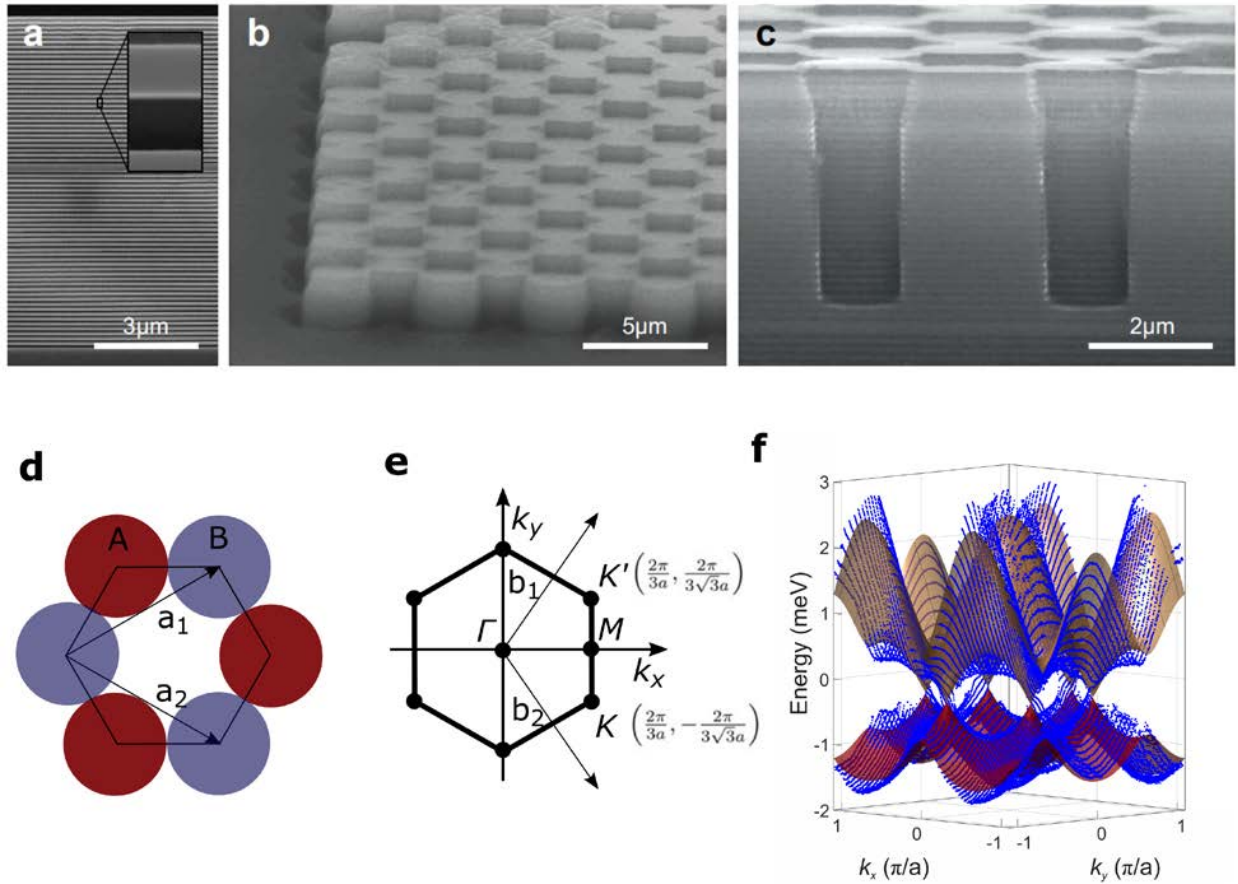

**Figure 2 | Lattice device layout and geometry.** (**a-c**) Scanning electron microscope images of the processed polariton honeycomb lattice. (**a**) Cleaved cross section of the microcavity prior to processing. (**b**) Tilted view of the half-etched honeycomb lattice for pillars with a diameter of $d = 2.0$ μm and an overlap $v = a/d = 0.85$. (**c**) Cleaved cross-section after etching, showing that only the top DBR has been etched, with the etching process stopped three to four λ/4-layers before the cavity. (**d**) Schematic of the real-space honeycomb unit cell with the two-element basis. (**e**) First Brillouin zone of the honeycomb lattice, most prominently featuring the six K- and K'-point. (**f**) Measured Fourier-space energy resolved photoluminescence of the investigated lattice in $k_y$-direction, scanning in $k_x$-direction. The blue data points are fitted to the measured dispersions, accurately revealing the six Dirac cones at the K- and K'-points. In red and yellow a tight-binding model is plotted, agreeing very well with the measured data.

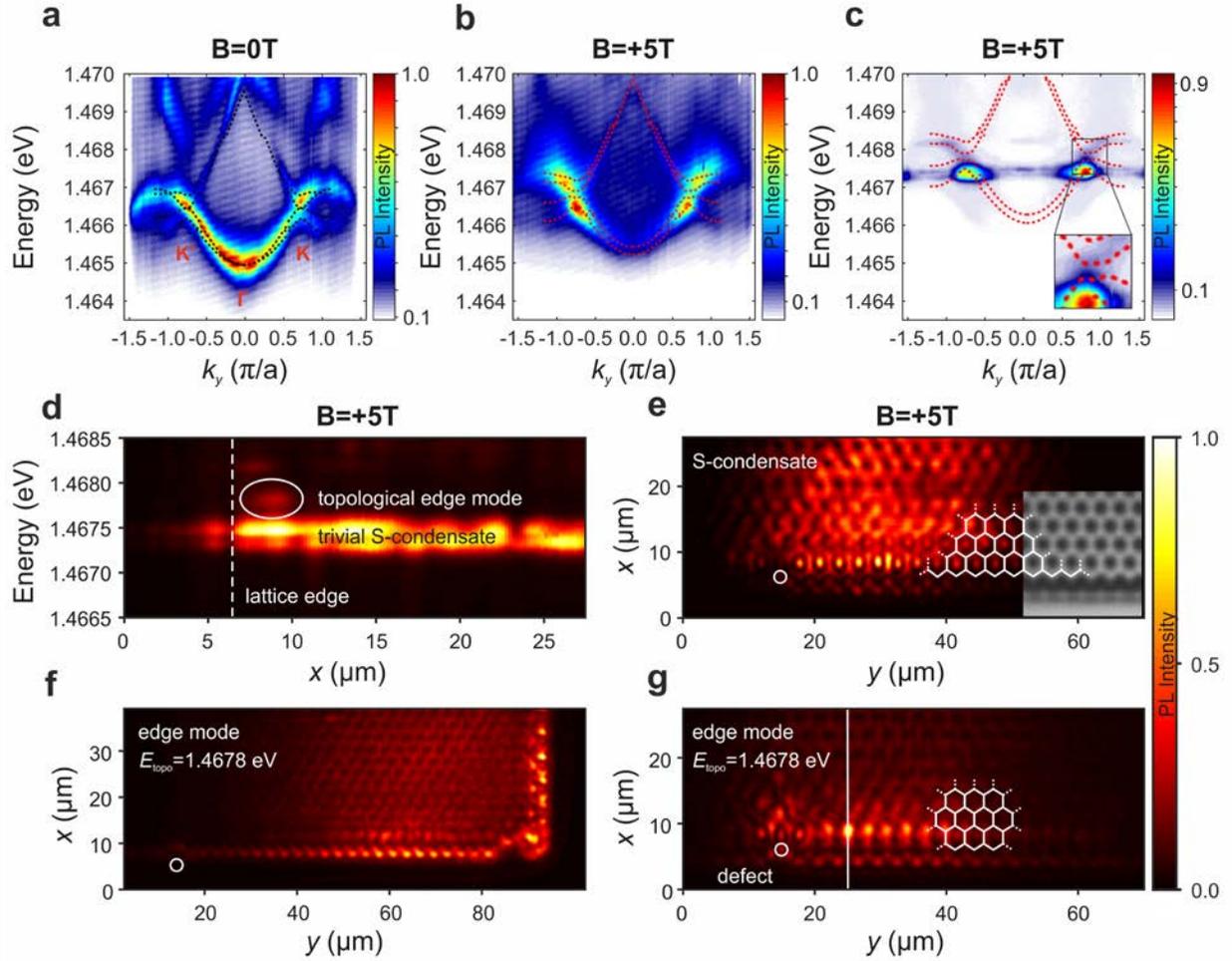

**Figure 3 | Photoluminescence measurements of a polariton condensate in a topological edge mode. (a)** Fourier space energy linear dispersion along $K' \to \Gamma \to K$ direction at $B = 0T$ compared with the calculated Bloch mode model (dotted black lines). **(b, c)** Equivalent dispersions at $B = +5T$, below and above threshold, respectively. Above a threshold of $P_{thr} = 1.8$ mW, we observe condensation into the $K, K'$−points of the S-band. Bloch mode calculations show that a distinct gap appears, as indicated by a dotted red lines. **(d)** Real-space spectrum in x-direction perpendicular to the zigzag edge along the straight white line in **(g)**. The real space x-axis in Fig. 3d, e and g is the same. The dashed white line marks the physical edge of the lattice. A trivial S-band condensate can be observed throughout the structure. At $E = 1.4678$ eV we observe the appearance of a localized mode, well separated from the bulk and located at the zigzag edge. **(e)** Mode tomography displaying the topologically trivial S-band condensate at $E_S = 1.4673 - 1.4675$ eV. A relatively homogeneous condensate within the pump spot diameter of 40 μm is observed. The inset shows an electron microscopy image of the structure. **(f)** Mode tomography of the topological edge mode at $E_{edge} = 1.4678$ eV at the corner position of the sample, showing clearly that the mode extends around the corner from the zigzag to the armchair configuration. **(g)** Mode tomography of the same position as in **(e)** of the topological edge mode at $E_{edge} = 1.4678$ eV. The mode is well located at the edge at $x \sim 8$ μm and in excellent agreement with the Bloch mode calculations shown in Figs. 1 e, f, g.

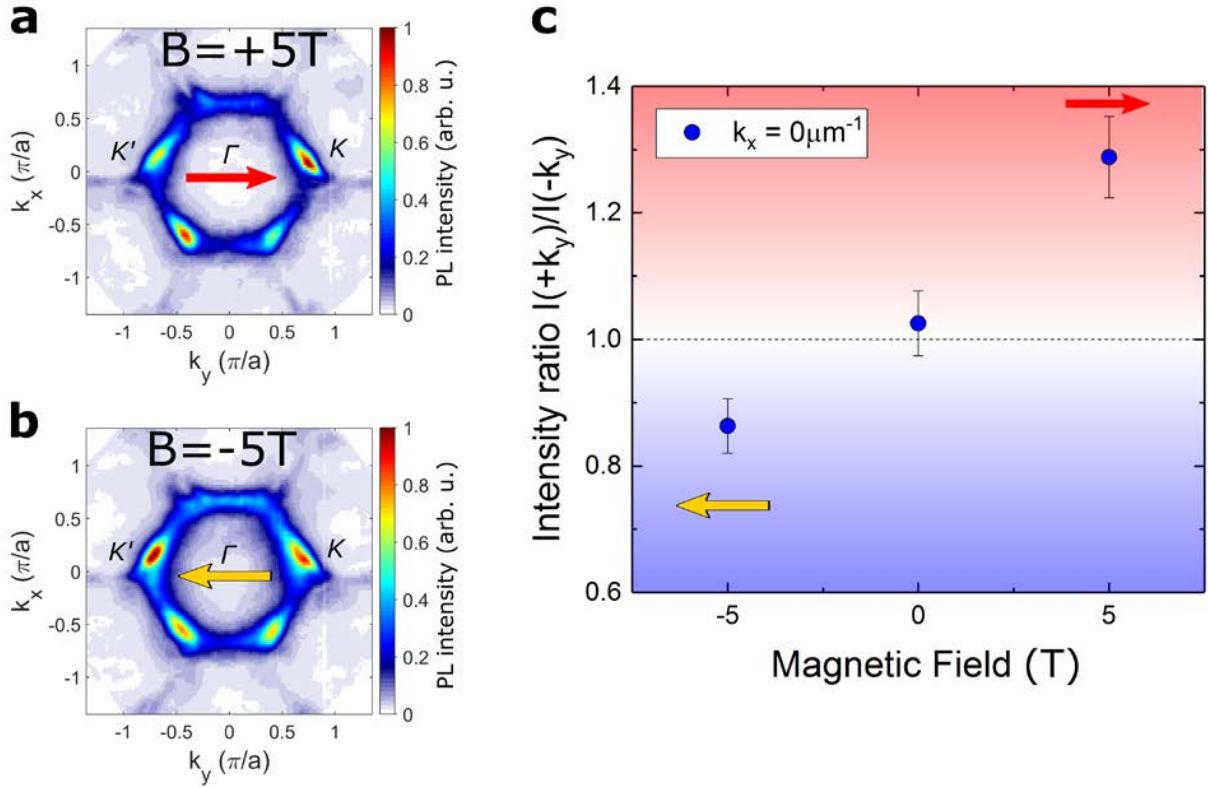

**Figure 4 | Chirality and propagation of the condensate.** Spectroscopic hyperspectral measurement of the S-band condensate at the K-points, including the topological edge modes for (**a**) $B = +5$ T, and (b) $B = -5$ T. The Fourier space is measured in $k_y$ and scanned in $k_x$. Subsequently, the full S-band (1.467 eV − 1.468 eV) including the energies linked to the edge state are displayed for $B = +5$ T (**a**) and $B = -5$ T (**b**). The zigzag edge is oriented in y-direction. (c) Polariton intensity ratio between the K'- and K-points in $k_y$-direction, for $k_x \simeq 0$ as a function of the applied magnetic field. One can clearly observe that the dominant propagation direction is inverted (yellow and red arrows) when the direction of the magnetic field is reversed, whereas for $B = 0$ T no dominant propagation direction along the edge is observed. The error bars in Fig. 4c originate from image distortions, inhomogeneities of the excitation, as well as uncertainties during the data processing, and are estimated at 5%.